# Title: Transitioning out of the Coronavirus Lockdown:
# A Framework for Zone-Based Social Distancing

**Authors:** Eric Friedman[1,2,3], John Friedman[2,3], Simon Johnson[3,4], Adam Landsberg[3,5]

**Affiliations:**

[1]International Computer Science Institute

[2]UC Berkeley

[3]COVID-19 Policy Alliance

[4]MIT

[5]Claremont McKenna College, Pitzer College, Scripps College

*Correspondence to: ejf@icsi.berkeley.edu or sjohnson@mit.edu

**Abstract:** In the face of elevated pandemic risk, is it necessary to completely lock down the population, imposing extreme social distancing? Canonical epidemiological models suggest this may be unavoidable for months at a time, despite the heavy social and human cost of physically isolating people. Alternatively, people could retreat into socially or economically defined defensive zones, with more interactions inside their zone than across zones. Starting from a complete lockdown, zones could facilitate responsible reopening of education, government, and firms, as a well-implemented structure can dramatically slow the diffusion of the disease. This paper provides a framework for understanding and evaluating the effectiveness of zones for social distancing.

**One Sentence Summary:** We present a simple epidemiological framework for understanding and evaluating the use of zones for social distancing in the face of a pandemic.

**Main Text:** As a result of its implications for health and mortality, the Coronavirus disease (COVID-19) pandemic has triggered massive disruptions to both economies and social structures.[1] In addition to more than 2 million infections and over 120,000 deaths worldwide by mid-April 2020, the resulting widespread lockdowns have depressed economic activity and sharply reduced the income of many people. In the United States, GDP will decline by at least 7 percent in the second quarter of 2020, and unemployment is expected by the Congressional Budget Office to exceed 10 percent.[2] The speed of this decline in measured economic activity is also dramatic: in one week, the number of new unemployment claims was ten times larger than in any single week of the 2007-08 recession.[3] By the second week of April 2020, unemployment in the United States was already around 13%, the highest rate since the Great Depression.[4]

---

[1] We use the official pandemic name from the World Health Organization, https://www.who.int/emergencies/diseases/novel-coronavirus-2019

[2] Latest data on infections and deaths: https://www.worldometers.info/coronavirus/?utm_campaign=homeAdvegas1?%20

[3] This is the CBO view on April 2, which may be considered low-ball estimate given where we are in mid-April: https://www.cbo.gov/publication/56314.

[4] This estimate is from the New York Times: https://www.nytimes.com/2020/04/03/upshot/coronavirus-jobless-rate-great-depression.html

The effect on human relationships is also unprecedented, with people effectively dissuaded from seeing friends in person, and actually forbidden (by force of law) from visiting loved ones in elder care facilities. The impact of lockdowns on individual health is also likely to prove significant.[5]

While the precise future course of infection is debated, many countries have begun to think about how best to transition out of the complete lockdown phase.[6] However, with good reason, the World Health Organization warns that abruptly ending any lockdown could result in new outbreaks, and the Centers for Disease Control and Prevention (CDC) is concerned there may be at least one more wave of infection "in the late fall, early winter where there will still be a substantial portion of Americans that are susceptible".[7] As a consequence, a number of prominent analysts have emphasized that social distancing should be relaxed in various gradual ways and on April 16, 2020, guidance issued by the White House seemed to support that general notion – while placing the onus on governors to decide the details.[8] In this paper, we provide a simple framework for designing and assessing different degrees of restriction on how individuals associate with others in person during a pandemic.

Our simple epidemiological model of "zone-based social distancing" offers a framework for more precisely estimating the efficacy of alternative social distancing measures. We develop a simple SIR epidemic model on a structured network, for which we can compute the inter-zonal reproduction number that can be used to guide further empirical analysis and decision making. While specific circumstances will always matter a great deal, the model suggests general advantages of organizing people into zones (i.e., a particular structure of groups), specifically so there can be strong interactions within each zone but weak interactions across zones.

Using this model, we evaluate the effect of various zone-based lockdown strategies combined with testing and mitigation measures. We then analyze several potential scenarios.

**Zones**: Zone-based social distancing works by dividing the population into groups ("zones"), so that interactions within a group are significantly greater than interactions between groups. Once a zone (group of people) becomes infected, it is completely isolated from other zones.

For example, one simple and potentially effective zone-based approach would be to allow people to go to work and interact in person as necessary in a research lab, office, or factory, but to limit

---

[5] https://www.sciencedaily.com/releases/2020/04/200408102137.htm
[6] "acceleration phase" is the CDC terminology as of April 7: https://www.cdc.gov/coronavirus/2019-ncov/cases-updates/summary.html.
[7] Head of the WHO: "Lifting restrictions too quickly could lead to a deadly resurgence," https://inews.co.uk/news/coronavirus-who-lifting-restrictions-too-early-dangerous-resurgence-2536303. Head of the U.S. CDC, "CDC Director says there may be another coronavirus wave in late fall an a `substantial portion of Americans' will be susceptible," https://www.newsweek.com/cdc-director-coronavirus-wave-late-fall-substantial-portion-americans-will-susceptible-1495401
[8] The White House guidance: https://www.washingtonpost.com/context/president-trump-s-proposed-guidelines-for-relaxing-social-distancing-guidance/b7768600-7906-408c-89a3-06fd5a47aa26/?itid=lk_inline_manual_3. Both ends of the American policy spectrum have detailed plans: https://www.aei.org/research-products/report/national-coronavirus-response-a-road-map-to-reopening/; https://www.americanprogress.org/issues/healthcare/news/2020/04/03/482613/national-state-plan-end-coronavirus-crisis/

non-work interactions to their own household. A version of this has been enacted in parts of Germany.[9]

Further compartmentalizing within companies can increase resilience. For example, one leading U.S. hospital has divided vital employees into one-week-on/two-weeks-off shifts. While working, an employee should self-isolate as much as possible, and, when returning to work, the employee is re-tested again. A prominent financial institution has two color teams, which are forbidden from meeting in person in any context.[10]

Another specific form of zone-based social distancing involves isolating towns or cities from each other, while allowing interactions within them, as in recent Italian travel restrictions.[11] Historically, the development of quarantine measures followed a similar heuristic: the entry of potentially infected people into an uninfected place was deliberately slowed down. During the Great Plague of 1665-1666, for example, "All trade with London and other plague towns was stopped."[12]

An issue of vital importance pertains to senior living facilities, where people are physically separated by living in different buildings or on separate floors within a building. In this case, transmission of disease across residents may occur as workers move about, with devastating consequences. A zone-based defense would specify that all residents and associated workers should stay within their specified zone – a policy that has significant implications, for example, for staffing levels during a pandemic. If some people, such as specialized medical professionals, must move between zones, then scarce testing and personal protective equipment (PPE) should be deployed so as to prioritize preventing inadvertent transmission.

Note that while zones structured around employment are important for economic activity, other zones could be designed to improve social and mental well-being. For example, older citizens who are self-isolating could be allowed to interact in small groups at a senior activity center, and families with young children who are isolating could interact with each other or with dedicated (intra-zone) childcare providers, who could be relatives or even paid employees working with a small number of families (and not interacting in person with anyone else in a different zone).

Also, one might consider the use of zones in carefully reopening schools and universities. For younger age groups, including elementary schools, zones could be based on shared homerooms, although there are obvious complications when a family has children who are in different grades. Taking these cross-connections into account can be quite complicated and must be done with great care – while also recognizing that prolonged loss of schooling has a likely serious negative and persistent effect on cognitive development, particularly in under-resourced environments where distance learning is not an effective option.

---

[9] Germany: public gatherings of more than two people (excluding people who live together and families) are not allowed. But going to work is allowed, although public transportation should be avoided if possible. https://www.studying-in-germany.org/social-distancing-guide-for-students-in-germany/. Bavaria and Saarland are under complete lockdown in mid-April 2020: https://www.dw.com/en/coronavirus-what-are-the-lockdown-measures-across-europe/a-52905137

[10] Hospital and financial institution cases are proprietary commercial information, which the authors have confirmed with at least two sources.

[11] Northern Italy under a travel ban that was later extended across the country, https://www.nytimes.com/2020/03/09/world/europe/italy-lockdown-coronavirus.html

[12] https://www.nationalarchives.gov.uk/education/resources/great-plague/

At universities, the natural zones could be based on living groups (e.g., dorms), which may be organized by academic area (e.g., undergraduate major) to better align with classes. This might allow in-person attendance for academic area specific lab and discussion courses while moving larger general education classes, which mix students from different fields, online.

This list is not meant to be exhaustive as many other zone-based strategies can and should be considered, based on the objective. The key point is that any movement of nonimmune people across zonal arrangements (e.g., university students going home for the summer), should be accompanied by comprehensive testing and appropriate transitory quarantine periods – based crucially on the probability of false negatives from these tests. Such decisions are likely to depend on specific relevant conditions and local or national social priorities, so the goal of this paper is to provide tools to allow such decisions to be analyzed, weighing economic and social benefits of such policies.

Our network analysis suggests that reducing lockdowns in a non-zonal fashion does not have the same effectiveness. Lockdown policies not based on zones may satisfy the important goals of reducing $R_0$ (the infectiousness of the disease) and protecting vulnerable populations, but such policies do not provide the significant additional protections offered by zone-based structures.

Note that our framework is not meant to replace standard approaches to reducing $R_0$, such as personal protective equipment, including facemasks, and social distancing in public spaces. Well-designed and properly implemented zone-based groups should be used to complement such approaches by providing public health officials, university administrators, company executives, and others with additional tools.

**Model**: We now present an informal overview of our model and the main results. The formal analysis is in the technical appendix.

We consider a standard stochastic Susceptible-Infected-Removed (SIR) network model.[13] That model considers a random network with n nodes and an average of d neighbors for each node. It starts with most nodes in the S (susceptible) state with the remaining nodes in the I (infected) state. Then in each period each infected node infects each of its neighbors with probability b and is removed (R) with probability g. This implies that a node remains infected for an average of T=1/g time periods before it transitions into the R (removed) state, which can correspond to recovery with immunity or mortality. The total probability that such a node infects its neighbor over that time period is q=b/(b+g).

A central established idea in the theory and practice of epidemiology is that the behavior of any epidemic depends crucially on the basic reproduction number, which in this model is given by $R_0$=q(d-1), i.e., the expected number of people that an I node infects.[14] Essentially, if $R_0$ <1 then the epidemic will typically die out after infecting only a small number of nodes while if $R_0$>1 then the epidemic is likely to spread widely, infecting a significant fraction of the population.

We now extend the standard model to an SIR model on a random 'zone-based network.' This network is divided up into m zones, where each zone contains n/m nodes. We write the average neighbors of a given node as $d = d_i + d_o$, where $d_i$ is the number of internal connections the node has to other nodes in its own zone and $d_o$ denotes refers to its outside connections to nodes in other zones. As we show in the technical appendix, one key parameter is the "inter-zonal connectivity ratio" ($C_R$) which is the fraction of a node's neighbors that are outside that node's own zone, i.e., $C_R = d_o/d$. Thus $C_R=0$ corresponds to a fully zoned (non-overlapping) social structure while $C_R=1$ correspond to a complete lack of effective zoning.

---

[13] Newman, M. E. (2002), Spread of epidemic disease on networks, Physical review E, 66(1), 016128
[14] Note that the formula for $R_0$ is for the case where all nodes have the same number of neighbors. This does not capture the existence of highly connected people and super-spreaders. https://www.nytimes.com/2020/04/12/us/coronavirus-biogen-boston-superspreader.html. In that case, the formula for $R_0$ is slightly altered, but the general threshold result is unchanged.

Given a zone-based social structure, we consider zone-based lockdown procedures. The key idea is that if we determine that there are infected nodes in a zone, then that zone is locked down – in the sense of having no further interactions with other zones. The goal of such a lockdown procedure is to minimize the number of zones which will become infected and need to be locked down. One key parameter in our analysis, which we call the "external infectivity time," is based on the number of infected people and the amount of time they were infected before lockdown was initiated. Suppose there were k infected people and the i'th person was infectious for time $t_i$ before lockdown, then we can define the "external infectivity time" $I_T = (t_1+t_2+…t_k)/T$, where we remind the reader that T is the expected duration of an infection for a particular individual.[15]

Analogous to the basic reproduction number $R_0$, one can construct an "inter-zonal reproduction number" $R_Z$, which is the expected number of other zones that will become infected from all the I nodes in a given zone. Similar to the basic reproduction number, $R_Z$ indicates the number of zones that will need to be locked down. If $R_Z <1$, then it is likely that only a small number of the zones will be infected and need to be locked down, while if $R_Z>1$ then it is likely that a large fraction of the zones will need to be locked down.

Thus, analogous to using $R_0$ to predict the spread of a disease, we can use $R_Z$ to evaluate zone-based policies. In particular, for a good zonal policy, $R_Z$ will be significantly smaller than $R_0$, and ideally less than one.

Using our model, we see that

$$R_Z = q(d_o-1)$$

Note that zone-based social distancing has 2 effects. First, as a form of lockdown, it reduces the number of interactions people have with others. Let $R_0$ be the reproduction number with no lockdown, and then let $R_N<R_0$ be the reproduction number where the reduction in d but not the effect of zoning is taken into account. Second is the effect due to the imposed zone-based network structure which reduces it further from $R_N$ to

$$R_Z = R_N*C_R*I_T$$

**The Value of Non-Overlapping Zones:** To understand the value of zone-based social distancing, and the pitfalls that may arise if it is done incorrectly, we consider the following location-based policies.

First, recall that if there is no lockdown then a reasonable estimate is $R_0=3$. Next, consider a partial lockdown in which people are allowed to socialize within a specified distance of their homes. In this case we do not get significant benefits from these overlapping zones as there is a large amount of cross contagion between zones. To quantify this, assume that this partial lockdown halves the number of interactions. Then one would expect $R_N=R_0/2=1.5$, since locking down is difficult to do effectively. However, if we had the same number of interactions with completely non-overlapping zones, such as made possible by locking down specific geographic regions, then this policy would be much more effective and $R_Z = R_N*C_R*I_T$. Under reasonable assumptions, $I_T=1.2$ is a plausible value, this zone-based policy could reduce the reproduction number by $C_R$, the fraction of interzonal interactions, which could easily be $C_R=1/5$, yielding $R_Z=.36$, a significant reduction.[16]

To see this more clearly, consider the distance-based lockdown in which people are allowed to socialize within 1 mile of their home (in the taxicab metric), and compare that with a zone-based lockdown with square zones of the same size. If $C_R=0$ then the zone-based lockdown will function perfectly, and no infection will ever cross the borders. It is easy to see that this is not true with the distance-based lockdown. For example, the most obvious strategy is to lockdown all homes within 1 mile of the detected infection, but it is easy to see that a small chain of infections can easily escape that region. In fact, for any region chosen there is some chance of the infection spreading beyond it and defeating the lockdown.

When $C_R>0$, the analysis is more complicated, but it seems intuitive that the zone-based lockdown will be better than the distance-based lockdown. To test this, we simulated this scenario directly on the 2-dimensional integer lattice with $d_i=950$, $d_o=50$, $q=0.5/(d_i+d_o)$, and a distance of 100. Thus $C_R=0.02$. We assumed that the time from infection to infectious was 2 days and the time from infection to symptomatic was 7 days. For these parameters the distance-based social distancing fails to contain the virus about 27% of the time while the zone-based social distancing only fails about 6% of the time. We need to enlarge the lockdown region by a factor of about 2.5 in order for the distance-based lockdown to achieve the same performance as the zone-based lockdown according to this metric.

**Workplace Zones:** One natural goal for zone-based defense is to re-open businesses without adversely affecting mortality rates. More broadly, interacting socially with exactly the same zone of people as seen (in person) at work could also be allowed. In this case, the internal connections would be between employees in the same company or division, while the most important external connections would arise through workers' housemates, who work in different companies (i.e., for other employers). One important insight here is that if a company discovers a single infected employee, then the company could immediately lock down, perhaps by requiring all their employees to quarantine at home.

However, since people who live together are highly likely to transmit the virus to each other, in the event of a zone becoming infected, all housemates of workers from that company should be immediately quarantined (and tested if possible) to prevent spreading to other companies.

**Nursing Homes as Zones:** Consider a group of nursing homes. Since much of the transmission is via staff members and specialists. In this case we can consider the network nodes to be the residents and consider the edges as determined by the staff interacting with residents, i.e. if we draw an edge between two residents if a staff member interacts with both. (We focus on the residents due to their extreme risk if infected.) This highlights the importance of regular testing of staff, which is a current issue of extreme importance but also – perhaps surprisingly – some controversy. It also allows us to quantify the risks associated with cross contamination between facilities, as these determine the Inter-Zonal Connectivity Ratio, $C_R$.

---

[16] See the technical appendix for details.

**Pooled Testing:** Pooled testing is when a single test is applied to a group of individuals and has begun to be applied to testing for COVID-19.[17] This test can then tell whether or not there are any cases in a group, but not who they are or how many have the virus. Pooled testing has the potential to dramatically improve the efficiency of zone-based social distancing, since we can pool an entire zone, dramatically reducing the number of tests required. For example, for the long-term care facilities in a particular state, there are not enough available coronavirus PCR tests. With pooled testing, it would be possible to pool all the residents and staff in a facility, and then only do further tests if there are any infections at that facility.

**Dynamic Implications:** One could add a dynamic element to these models by assuming random infections springing up in zones, arising from interactions outside our main model. One could use these to "flatten the curve" in a controlled dynamic fashion. For example, when hospitals are reaching capacity, a planner could pre-emptively lockdown some of the zones, thereby reducing the number of potential new infections. This would also lower $R_Z$, hence reducing the probability of a large epidemic. Alternatively, as the number of cases ebbs, one could merge zones – allowing more interactions – or allow more interactions between zones, increasing $C_R$.

**Conclusions:** Zone-based policies could be used effectively in the transition out of wide-scale lockdowns, and further calculating the zonal reproduction number ($R_Z$) will help develop and evaluate such policies.

---

[17] https://www.technion.ac.il/en/2020/03/pooling-method-for-accelerated-testing-of-covid-19/,https://www.ketv.com/article/nebraska-public-health-lab-begins-pool-testing-covid-19-samples/31934880#

**Technical Appendix:**

**Model and Results:** In this section we present the formal analysis of our model. We first consider the SIR model on a random d-regular network with n nodes, i.e. all nodes have degree d. The restriction to fixed d is meant to simplify the presentation. Alternative models with different degree distributions behave similarly. We assume that we start with most nodes in the S (susceptible) state and a small number in the I (infected) state. In each period an infected node will infect its susceptible neighbor with probability b and become R (removed) with probability g, which implies that the average time in an infectious state is $T=1/g$. This implies that the transmissibility $q=b/(b+g)$ which is the probability that an infected node infects its neighbor overall. For large n and d the following result is known:

*Result:*[18] Let $R_0 = q(d-1)$ then if $R_0<1$, the number of infected nodes is less than $1/(1-R_0)$ while if $R_0>1$ the expected number of infected nodes is given by $nf(R_0)$ where $f(R_0)$ is an increasing function of $R_0$ which approaches 1 as $R_0$ approaches infinity.

Next consider an SIR model on a random zonal network. We assume there are m zones, which are groups of nodes, each of size $s=n/m$. Now we assume that each node has $d_i$ edges to nodes in its zones and $d_o$ to nodes in other zones, where $d=d_i+d_o$. Define $D_R = d_o/(d-1)$. We now consider the zone meta-network where each node is an entire zone of the original network. Thus, the zone meta-network has m nodes and we will apply the above result to it.

Assume a zonal lockdown procedure Given a lockdown with k infected people where the i'th person was infectious for time $t_i$ before lockdown, define the effected external infectivity time to be $I_T = (t_1+t_2+\ldots t_k)/T$. Now, consider any monotonic limit where both m and n/m both approach infinity. The dynamics on the zonal meta-network behaves like an SIR network with degree $d_Z=sd_o$ and transmission probability $I_Tq/s$.

Combining all of these we see that the basic reproduction number of the zone meta-network,

$R_Z = R_0*D_R*I_T$.

Thus, the zonal reproduction satisfies the above proposition for the zone meta-network.

Note that in the main text of this paper we approximate $D_R$ by $C_R = d_o/d$ since it is easier to use and very close numerically in most situations.

Also, note that one can easily generalize this result to include different random graph models including arbitrary degree distributions in which case the formulas for $R_0$ changes slightly but is usually well approximated by $q_d$, where d is the average degree of a node.

**Estimating $I_T$:** Consider a policy where we test for people that show symptoms of COVID-19 and lock down a zone if any person in that zone tests positive for the virus. Assume that the person becomes infectious after 2 days and symptomatic in 7 days when the zone will be locked down. Suppose in day 3 that person infects 1 person and in day 4 that person infects another. Then in day 5 the first person he infected becomes contagious and infects one more person, on day 6 they infect 3 people between them. Then $t_1=5$, $t_2=4$, $t_3=3$, $t_4=2$, $t_5=t_6=t_7=1$. Now,

---

[18] This result is found in Newman, M. E. (2002), Spread of epidemic disease on networks, Physical review E, 66(1), 016128 and Kenah, E., & Robins, J. M. (2007), Second look at the spread of epidemics on networks, Physical Review E, 76(3), 036113.

$I_T=17/14=1.2$, where we used T=14 as a reasonable value based on WHO estimates for mild cases.[19]

---

[19] https://www.who.int/docs/default-source/coronaviruse/who-china-joint-mission-on-covid-19-final-report.pdf p.14